\documentclass[conference]{IEEEtran}
\IEEEoverridecommandlockouts
\usepackage{cite}
\usepackage{amsmath,amssymb,amsfonts}
\usepackage{algorithmic}
\usepackage{graphicx}
\usepackage{textcomp}
\usepackage{xcolor}

\ifCLASSOPTIONcompsoc
\usepackage[caption=false,font=normalsize,labelfon
t=sf,textfont=sf]{subfig}
\else
\usepackage[caption=false,font=footnotesize]{subfi
g}
\fi

\def\BibTeX{{\rm B\kern-.05em{\sc i\kern-.025em b}\kern-.08em
    T\kern-.1667em\lower.7ex\hbox{E}\kern-.125emX}}
\begin{document}

\title{Neuromorphic implementation of ECG anomaly detection using delay chains %
\thanks{This work was supported by the European Research Council (ERC)
under the European Union’s Horizon 2020 Research and
Innovation Program Grant Agreement No. 724295 (NeuroAgents).} } %

\author{\IEEEauthorblockN{Stefan Gerber\IEEEauthorrefmark{1}, Marc Steiner\IEEEauthorrefmark{1}, Maryada\IEEEauthorrefmark{1}, Giacomo Indiveri\IEEEauthorrefmark{1}, and
Elisa Donati\IEEEauthorrefmark{1}}
\IEEEauthorblockA{\IEEEauthorrefmark{1}Institute of Neuroinformatics, University of Zurich and ETH Zurich, Zurich, Switzerland\\ 
Email: stefanerik.gerber@uzh.ch, marc.steiner@ini.uzh.ch, maryada@ini.uzh.ch, giacomo@ini.uzh.ch, elisa@ini.uzh.ch}
}

\maketitle

\begin{abstract}
  Real-time analysis and classification of bio-signals measured using wearable devices is computationally costly and requires dedicated low-power hardware.
One promising approach is to use spiking neural networks implemented using in-memory computing architectures and neuromorphic electronic circuits.
However, as these circuits process data in streaming mode without the possibility of storing it in external buffers, a major challenge lies in the processing of spatio-temporal signals that last longer than the time constants present in the network synapses and neurons.
Here we propose to extend the memory capacity of a spiking neural network by using parallel delay chains.
We show that it is possible to map temporal signals of multiple seconds into spiking activity distributed across multiple neurons which have time constants of few milliseconds.
We validate this approach on an ECG anomaly detection task and present experimental results that demonstrate how temporal information is properly preserved in the network activity.
\end{abstract}

\section{Introduction}
\label{sec:intro}
The recent advances in wearable device technologies enable constant monitoring of people's vital signs, such as with electromyography (EMG) or electrocardiogram (ECG) signals~\cite{Donati_etal19, Ramasamy2018WearableReview}.
In recent years significant progress has been made in automatically classifying ECG signals, mainly using ANNs~\cite{Mathews2018AClassification, Wang2019Energy-EfficientDevices, Sannino2018ADetection, Kiranyaz2016Real-TimeNetworks}.
However, storing and then transmitting this data for off-line classification comes with high energy costs.
Processing and classifying signals in real-time directly on-site can greatly reduce the amount of data that has to be stored and transferred and therefore lead to a significant reduction in energy consumption.
Inspired by biological neural networks, spiking neural networks (SNN) have been proposed as a possible solution to reduce the power cost~\cite{Amirshahi2021ECGDevices, Das2018UnsupervisedReadout} since they operate in an event-driven, sparsely activated domain~\cite{Rueckauer2017ConversionClassification}. Further decrease in power consumption can be obtained by using analog sub-threshold circuit neuromorphic systems, where the dynamics are matched to the physical properties of neurons and synapses~\cite{Mead1990NeuromorphicSystems, Indiveri2011NeuromorphicCircuits}. 
The neuromorphic approach supports the implementation of networks that are radically different from those used in classical von Neumann architectures due to how memory and processing are organized. In neuromorphic architectures, the memory is distributed with the processing decreasing the required computational power while
hindering the possibility to store information to process in batch as in classic machine learning~\cite{Indiveri2015MemorySystems}. Hardware implementations of recurrent SNNs on neuromorphic chips for anomaly detection in multichannel ECG signals have been recently proposed~\cite{Bauer2019Real-TimeProcessor, Corradi2019ECG-basedHardware}. However, it is difficult to try and interpret recurrent SNNs, and they often require fine-tuning of the parameters, which is challenging on neuromorphic hardware. 
Here we propose a simpler approach by implementing a delay chain network in neuromorphic hardware that manages to capture temporal data in space and is well-suited for classification networks down the line. This approach is flexible and extendable and can therefore be tuned to work for different kinds of biomedical signals that require working memory over different timeframes and a variety of temporal resolutions.

\section{Methods}
\label{sec:imethods}
\subsection{Neuromorphic Hardware}
\label{ssec:NH}
To validate our model we used the DYNAP-SE (Dynamic Neuromorphic Asynchronous Processor) chip~\cite{Moradi2018ADYNAPs}. This device comprises analog subthreshold circuits that emulate neuron and synapse dynamics, as well as asynchronous digital circuits to manage address event traffic. The neurons are based on a model of the Adaptive Exponential Integrate and Fire neuron~\cite{Brette2005AdaptiveActivity}. The synapses exhibit first order low-pass filter dynamics using a Differential Pair Integrator circuit~\cite{Bartolozzi2007SynapticVLSI}. The chip integrates four cores of 256 neurons each, allowing the implementation of real-time spiking neural networks. The neurons have 64 inputs each, which can be configured to use one of four possible synapse types (slow/fast excitatory and subtractive/shunting inhibitory). This leads to a vast array of different possibilities for practical applications, that require real-time processing of event-based sensory signals and ultra-low power. It also makes it well suited to use in neural processing systems that can be directly interfaced with biomedical sensors for on-line classification. 

\subsection{ECG signals and conversion to spikes} \label{ecgSignals}
\begin{figure}[htbp]
\centerline{\includegraphics[width=0.45\textwidth,keepaspectratio]{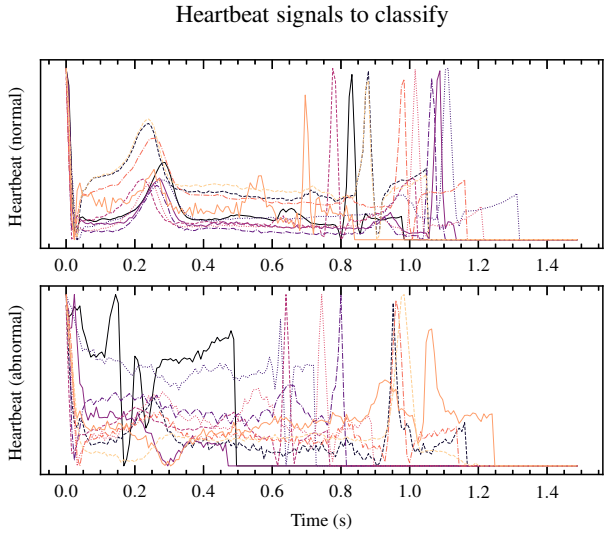}}
\caption{ECG signals used as input for the network and subsequent classification. Only a subset of 10 signals is shown for better visualization.}
\label{fig:ecgSignals}
\end{figure}

As ECG dataset we used the open-access database Physionet  (PTB Diagnostic ECG database~\cite{Bousseljot1995NutzungInternet} and MIT-BIH Arrhythmia Dataset~\cite{Moody2001TheDatabase}). The PTB dataset consists of ECG signals from 12 leads of patients with myocardial infarction vs healthy controls sampled at 1000Hz. The MIT-BIH dataset consists of beat type annotations made by at least two physicians. The signals are sampled at a rate of 360Hz.
Already preprocessed signals from~\cite{Kachuee2018ECGRepresentation} were used. The annotated beat types have been mapped to 5 categories according to AAMI EC57 standards, grouping different kinds of anomalies. Fig. \ref{fig:ecgSignals} shows an example of preprocessed and labeled ECG signals. The signals have been segmented, down sampled to 125Hz, and padded with zeros to make them equal length. Only ECG lead II was used in both datasets.

To convert the signal into spikes we simulated multiple Analog-to-Digital Delta Modulators (ADMs) with different thresholds~\cite{Sharifshazileh2021AnEEG}. Each ADM produces two spike trains: every time the input signal changes by a positive (negative) amount equal to the threshold parameter, the ADM will generate one spike for the UP (DOWN) channel. We converted the input signals with three different ADMs producing 6 different spike trains (3 pairs of UP and DOWN channels) as input to the chips spike generator. We set 3 threshold parameters such that one pair generates spikes only when there are big changes in the signal (threshold of 0.2), one for medium changes (threshold of 0.1), and one for small changes (threshold of 0.04).

\subsection{Network architecture} \label{ssec:network}
Short working memory was achieved by connecting neurons in a chain, and leveraging their low pass filter properties to create maximal delay from the spike event arriving at the neuron to the generation of the next event for the following neuron in the chain. As the spike pattern propagates slowly through the chain, a neuron further down in the chain has the same activity as a neuron before had a short time ago, as long as every neuron exactly replicates its input spikes. To keep the spike pattern consistent across the whole chain, the parameters of the neurons were tuned such that every input spike would generate exactly one output spike. Further parameters were adjusted, in favor of high neuron time constants, such that the delay was as long as possible for each individual neuron. Since the neurons tuned like this are very slow to react, the behavior of recreating the input spike pattern would only work for low input frequencies of up to 50 Hz (depending on the specific neuron).

\begin{figure}[htbp]
\centerline{\includegraphics[width=0.45\textwidth,keepaspectratio]{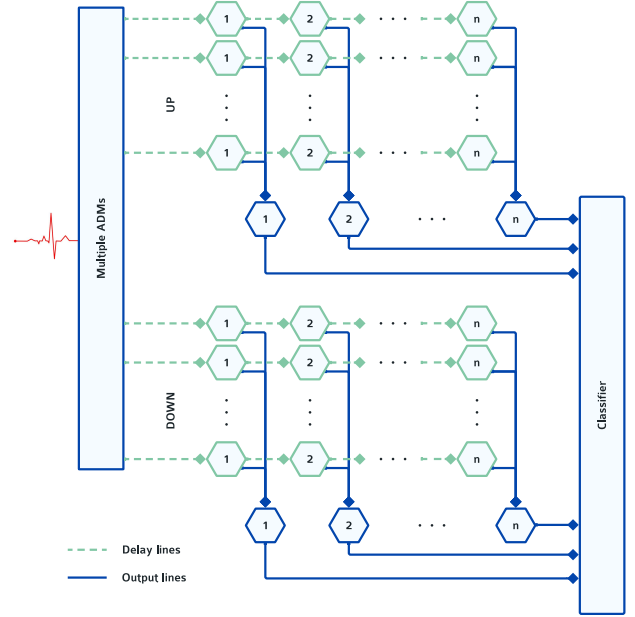}}
\caption{Network architecture: The ADM processes the signal and outputs UP and DOWN spikes, which are then fed into the parallel delay chains. The output neurons then integrate the contributions of each delay chain for different steps. Each output neuron represents the input signal at a different point in time. The activities of these neurons can then be used simultaneously as input features for a classifier.}
\label{fig:networkArchitecture}
\end{figure}

Normally ADMs would need parameters that generate spike patterns with way higher frequencies than this to not lose too much information in the conversion process. This is why we resorted to using three different ADMs. Each channel from the ADMs is connected to its own delay chain. This creates a network of parallel delay chains that all carry a different part of the information in the original signal (see Fig.~\ref{fig:networkArchitecture}). Because of device mismatch, not every neuron behaves the same and has the same delay. We tackled the variability of neurons and synapses by individually selecting ones with desired properties. To not distort the input signal the delay of individual neurons was measured, during a calibration phase, and the network was set up in such a way that at every step in the chain~\cite{Sheik2012ExploitingDelays}, all neurons from the parallel chains had a delay that was as similar as possible. This makes sure that no spike train in one of the chains travels faster than one in another and the information is kept in sync. 

The neurons from all chains at every step corresponding to UP channels were connected to one fast output neuron. The same was done for all chains corresponding to DOWN channels. Different weights were used for these connections depending on the threshold of the ADM the chain originates from. The result of this is that spikes generated in ADMs with higher thresholds will result in a higher firing frequency in the output neurons than spikes coming from ADMs with lower thresholds. The exact behavior is shown in Fig.~\ref{fig:outputNeurons}.

This setup allows using ADMs with relatively big thresholds, that encode the signal in sparse spike patterns. The sparse spike patterns make it possible to propagate the signal with slow neurons without losing the exact pattern and therefore create chains with long delays that implement a working memory with the longest duration possible. The total information is then accessible in form of the activity of the output neurons, which integrate back all the sparse spike trains into a faithful representation of the original analog signal.

The proposed architecture can be composed of more or fewer lines, consisting of the desired number of neurons, to match the required task. The only limit is presented by the number of neurons on the chip. In the proposed application, we deployed one core for the delay neurons and one ore for the output neurons, such that we could tune neurons and synapses parameters differently.

\subsection{Classification} \label{sec:class}
After setting up a full delay network as shown in Fig.~\ref{fig:networkArchitecture}, consisting of 6 delay chains with 15 neurons each as well as 15 output neurons for the UP and DOWN channels, we evaluated its performance by using the aforementioned heartbeat signals. For both datasets, 1000 signals were randomly sampled. It was made sure that all labels were represented equally often. The signals were converted into the 6 spike trains using the 3 ADMs and these spike trains fed into the network. The mean firing rate over 30ms of the output neurons was measured after 0.45s, 0.9s, and 1.35s. We had to measure three times because after 0.45s the first part of the signal reaches the end of the delay chain (see Fig.~\ref{fig:delayNeurons}). 
The measured firing rates served as features to train a software classifier. 
To understand the performances of the proposed spiked-based feature extraction method, we compared the accuracy of this classifier with the accuracy of a classifier trained on the raw data.

\section{Results}
\subsection{Neuron properties}

\begin{figure}[htbp]
\centerline{\includegraphics[width=0.45\textwidth,keepaspectratio]{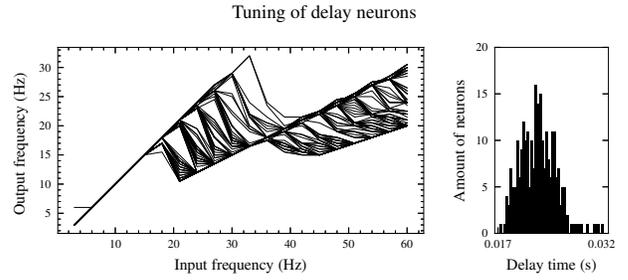}}
\caption{Input frequency to output frequency relationship and delay times for the delay neurons. Exhibiting a low one-to-one frequency range but in turn high delay times.}
\label{fig:delayNeurons}
\end{figure}

To preserve the input signal through the delay chains, having a linear one-to-one input-to-output frequency relationship is essential. At the same time, the neurons have to be very slow to reach a significant amount of delay without using many neurons. Using a lot of neurons per delay chain increases the probability of the signals in the different parallel chains getting out of sync, which leads to the summed-up information in the output neurons becoming meaningless. But if the neurons are too slow, they will not be able to keep a one-to-one firing rate for higher frequencies, and more parallel lines are needed. This trade-off between delay and maximal one-to-one firing frequency has to be optimized depending on the application. In this case, the neurons were tuned and selected for a delay of about $0.02s$ per neuron and a one-to-one frequency up to at least $20Hz$, as seen in Fig. \ref{fig:delayNeurons}.

\begin{figure}[htbp]
\centerline{\includegraphics{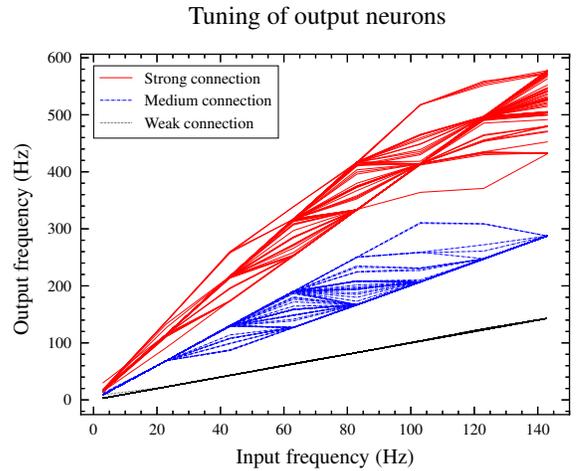}}
\caption{Input vs output frequency relationship for the output neurons. The firing rates for the same neurons are plotted three times, each time with different connection strengths of the input. By using a connection strength between delay chain neurons and output neurons that is proportional to the corresponding ADM threshold, one can accurately preserve the information in the input signal.}
\label{fig:outputNeurons}
\end{figure}

Output neurons on the other hand do not suffer from the restriction of having to be slow. They have to be fast and respond differently depending on which delay chain (and subsequently which ADMs) they receive a spike from, to faithfully reconstruct the inputs from all parallel chains (see Fig.~\ref{fig:outputNeurons}).

\subsection{Spike train is preserved through delay chains}

\begin{figure}[htbp]
\centerline{\includegraphics[width=0.45\textwidth,keepaspectratio]{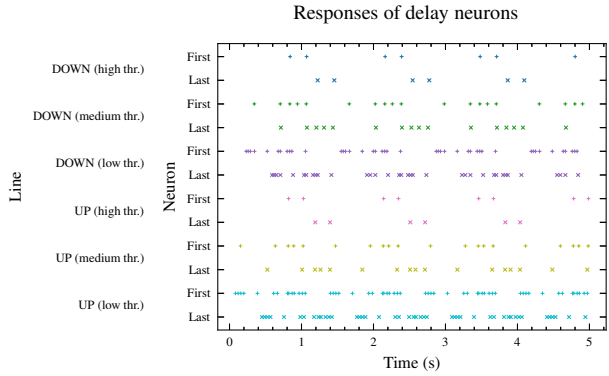}}
\caption{The responses of the first and last neurons in the 6  parallel delay chains are shown in response to a repeating heartbeat signal converted by the different ADMs. Neurons of the same chain are plotted in the same color. One can see that the input signals are preserved along each delay chain.}
\label{fig:spiketrainPreservation}
\end{figure}

Comparing the spike patterns of the first neuron and the last neuron of every chain shows how good the signal is preserved through each chain. A single repeating spike pattern generated from one heartbeat signal was distributed over the chains and the first neuron response and the last neuron response in each chain were captured. As shown in Fig.~\ref{fig:spiketrainPreservation}, besides a small amount of noise, the signal exits the chains almost exactly how it enters it, just with the expected delay. 

\begin{figure}[htbp]
\centerline{\includegraphics[width=0.45\textwidth,keepaspectratio]{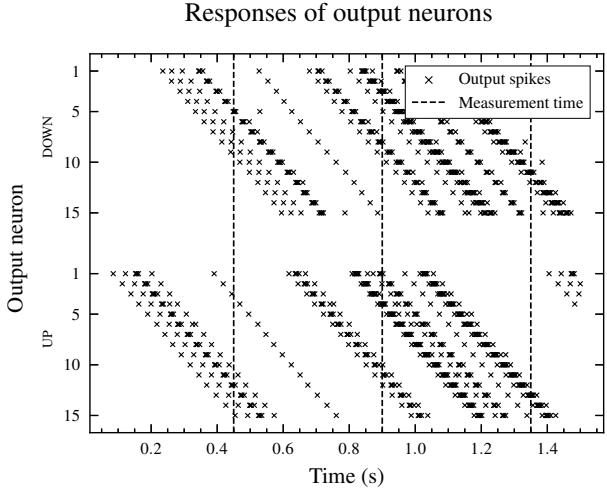}}
\caption{Delay chains consisting of 15 neurons each were used. Consequently, 30 output neurons were needed, 15 for the chains receiving the DOWN input signal (shown on top) and 15 for the chains receiving the UP signal (shown on the bottom). The vertical lines show multiples of the expected total delay of the network, which corresponds to the length of the stored signal. Hence after this time, the output has to be measured again otherwise part of the signal information is lost.}
\label{fig:signalPreservation}
\end{figure}

\subsection{Total signal is preserved throughout the output neurons}

The response of the UP and DOWN output neurons to the same signal is shown in Fig.~\ref{fig:signalPreservation}. The signal, besides a small amount of noise, stays consistent along the delay chains, with longer and longer delays for every consecutive output neuron.

\subsection{Temporal information converted into spatial neural activities}

\begin{figure}[htbp]
\centerline{\includegraphics[width=0.45\textwidth,keepaspectratio]{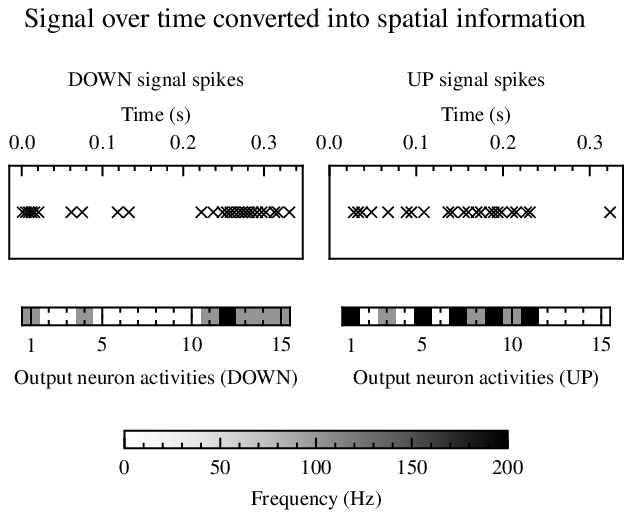}}
\caption{The input spikes from 3 ADMs with different thresholds are combined (once for DOWN and once for UP spikes) and compared to the corresponding UP and DOWN output neuron activities. One can see, that the information of the input signal over time is converted into neuron firing rates in space.}
\label{fig:signalInSpace}
\end{figure}

Another way to see how the information of the input signal is preserved over time is by looking at the firing rate of the output neurons after the signal reaches the end of the network (dashed lines in Fig.~\ref{fig:signalPreservation}). To get an estimate for the firing rate, we used a small sampling time of 0.03s. In Fig. \ref{fig:signalInSpace} the firing rates of the output neurons are shown for the DOWN and UP input signals respectively. We can see that the firing rates of the output neurons correspond to the input signals over time. Therefore the temporal information is converted into spatial neural activities, hence it acts as short-term working memory.

\subsection{Signal classification}

Having the ability to store temporal information in the short-term memory provided by these neurons, we can now investigate how accurate the information in the signal is preserved, by comparing the accuracy of the classification of output neuron activities with the classification done on raw data. For the PTB dataset, we reached an accuracy of 86.3\%, while the classification of the raw data got up to 90.0\% accuracy. For the MIT-BIH dataset, the five labels were classified with an accuracy of 82.0\%, while the classification on the raw data reached 85.6\%. This small drop in accuracy for both datasets is further evidence that most information in the signal can be preserved using our network. In theory, this classification step could also be implemented directly on-chip, hence allowing for the energy-efficient classification of signals in real-time.

\section{Conclusion}

Creating working memory in spiking neural networks is still an open challenge that is particularly difficult for mixed-signal hardware implementations, with limited precision and device mismatch. We show that using a rather simple feed-forward network that uses relatively few neurons, one can reliably convert a temporal spike pattern into neural activity in space, hence providing a short-term memory. This memory can then be used to classify biomedical signals such as ECG heartbeats. Because this approach is simple, modular, and comprehensible, the network can easily be extended for the use of signals requiring more delay, higher temporal resolution, or higher firing frequencies. This makes it an extremely useful building block at the beginning of classification tasks of biomedical signals using spiking neural networks.

\bibliography{references}
\bibliographystyle{IEEEtran}

\vspace{12pt}

\end{document}